# Electron spin polarization in strong-field ionization of Xenon atoms

Alexander Hartung[1,†], Felipe Morales[2], Maksim Kunitski[1], Kevin Henrichs[1], Alina Laucke[1], Martin Richter[1], Till Jahnke[1], Anton Kalinin[1], Markus Schöffler[1], Lothar Ph. H. Schmidt[1], Misha Ivanov[2], Olga Smirnova[2] and Reinhard Dörner[1]

[1] Institut für Kernphysik, Goethe Universität Frankfurt, Max-von-Laue-Straße 1, D-60438 Frankfurt, Germany.
[2] Max-Born-Institut, Max-Born-Straße 2A, D-12489 Berlin, Germany.
[†] Correspondence to: hartung@atom.uni-frankfurt.de

**Abstract**

As a fundamental property of the electron, the spin plays a decisive role in the electronic structure of matter from solids to molecules and atoms, e.g. causing magnetism. Yet, despite its importance, the spin dynamics of electrons released during the interaction of atoms with strong ultrashort laser pulses has remained unexplored. Here we report on the experimental detection of electron spin polarization by strong-field ionization of Xenon atoms and support our results by theoretical analysis. We found up to 30% spin polarization changing its sign with electron energy. This work opens the new dimension of spin to strong-field physics. It paves the way to production of sub-femtosecond spin polarized electron pulses with applications ranging from probing magnetic properties of matter at ultrafast time scales to testing chiral molecular systems with sub-femtosecond temporal and sub-Ångström spatial resolution.

**Main Text**

Short laser pulses provide an electric field which can be strong enough to suppress the binding potential of an atom or molecule and lead to field ionization. Electrons passing over the barrier, or tunneling just under it and emerging from the atom are subsequently driven by the laser field. So far, nearly all works exploring the electronic behavior after ionization have solely used the binding energy of the electron and the shape of the barrier as defining properties, omitting the other fundamental property of the electron – its spin. This is even more surprising as a few pioneering theoretical works have indicated the importance of the spin of the outgoing electron in strong field ionization[1,2]. From the fundamental quantum standpoint, the spin of the liberated electron should not be ignored since, first, there is correlation between this electron and the ion left behind, and, second, ionization is known to trigger spin-orbit dynamics in the ion[3]. Rare gas atoms with their closed shells and overall vanishing spin provide an ideal starting point for studies of such strong-field spin effects.

For single-photon ionization, the spin polarization of photoelectrons ejected from the outermost orbital has been thoroughly studied experimentally[4] and theoretically[5-8]. The physics behind spin polarization in this case however is completely different from the strong-field regime discussed here. In the single-photon case, photoelectrons of the same energy populate a small set of continuum angular momentum states, as dictated by the dipole selection rules. Different phase shifts for each such set of continuum states then lead to spin polarization of the energetically degenerate electrons[5]. The single-photon case was generalized to the weak-field multiphoton regime [6-8], uncovering the importance of intermediate resonances. In contrast, in the case of over-barrier strong-field ionization discussed here, electrons of different spin (but same binding energies) are substantially shifted in their kinetic energy in the continuum. The shift is due to different direction of their bound state momentum relative to the sense of rotation of the laser field and the correlation between the bound state momentum and the spin of the released electron, see Fig.2 and the discussion below.

For our present study we exposed Xenon atoms to circularly polarized 780 nm, 40 fs laser pulses. The peak intensity was estimated to be $I_0 \sim 3.3 \cdot 10^{14}$ W/cm$^2$ at the center of the focal spot. The intensity is well above the saturation intensity of Xe[9-10], leading to ionization already on the rising edge of the pulse, as expected in the saturation regime. The effective intensity $I_{eff}$ at which most of the electrons were released can be extracted from the peak position of our experimental electron energy distribution, which for circularly polarized pulses is around the electron ponderomotive energy $U_p = e^2F^2/2m_e\omega^2$ (F is the electric field amplitude, ω is the laser frequency, $e$ and $m_e$ are the electron charge and mass), see e.g. Ref. 11. This estimate yields $I_{eff} \approx 1.4 \cdot 10^{14}$ W/cm$^2$, consistent with field ionization with little or no tunneling. We measured the kinetic energy and spin polarization of the electrons with a time-of-flight spectrometer equipped with a commercial Mott polarimeter[12]. Figure 1 shows measured spin polarization as a function of the electron kinetic energy. Experimental results are in good agreement with our numerical simulations.

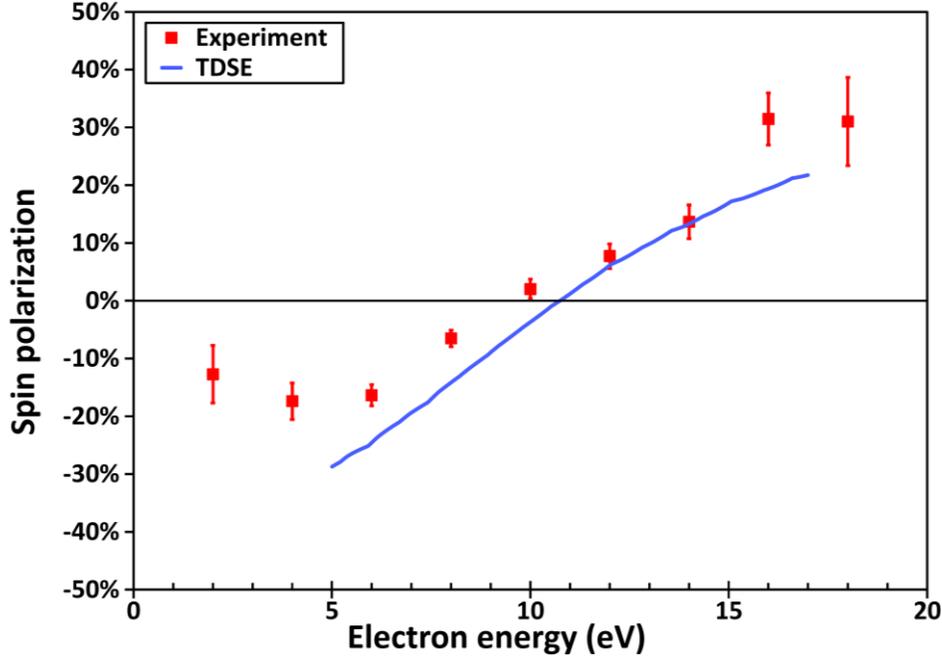

**Figure 1 | spin polarization of electrons ejected by strong field ionization of Xe parallel to the light propagation direction by circularly polarized laser pulses.** The spin polarization is defined as the weighted difference between spin-up and spin-down electrons, see Supplementary Material. Consequential positive values correspond to a surplus of electrons with spin parallel to the propagation axis of the laser. Red rectangles show experimental data for 40 fs, 780 nm pulses. Solid blue curve shows results of numerical simulations. Error bars show statistical errors only.

The basic physics behind our observation is explained in Fig. 2A, which shows an artist's view of the ionization process from the highest 5p j = 3/2 orbital of Xe. We chose the quantization axis to be the light propagation direction (orange), with positive projection of the total angular momentum $m_j$ being in light direction. The two 5p j = 3/2 orbitals with $m_j = +3/2$ (red) and $m_j = -3/2$ (blue) are degenerate in energy. For $m_j = 3/2$ ($m_j = -3/2$) the electron spin is oriented upwards (downwards) and the electrons in the $m_\ell = +1$ ($m_\ell = -1$) state rotate in the same (opposite) direction as the circularly polarized laser field. The electronic wave packet freed from these orbitals crosses the saddle of the potential with some initial momentum $p_{initial}$ tangential to the donut-shaped orbital and hence perpendicular to the direction of the laser electric field at that time. The wave packet is then driven by the circular laser field, which in the end results in a net momentum transfer of $p_{streak} = \sqrt{2 \cdot U_P}$. Depending on the sign of $m_\ell$ the initial momentum $p_{initial}$ is parallel or antiparallel to the streaking momentum, i.e. $p_{final} = p_{streak} \pm p_{initial}$. Therefore, the spectra of the photoelectrons ejected from these two $|m_j| = 3/2$ orbitals are offset in energy. Crucially, they also have an opposite direction of the spin. Indeed, the sign of $m_\ell$ and the spin state $m_s$ are uniquely linked for each $m_j$ orbital. Due to this link, the direction of electron spin is correlated to an increase or decrease of the kinetic energy of electrons emitted from strong field ionization in circularly polarized fields. Thus, electrons with low kinetic energy will have their spin pointing downwards while electrons with high energy will have their spin pointing upwards, in agreement with our results in Fig. 1.

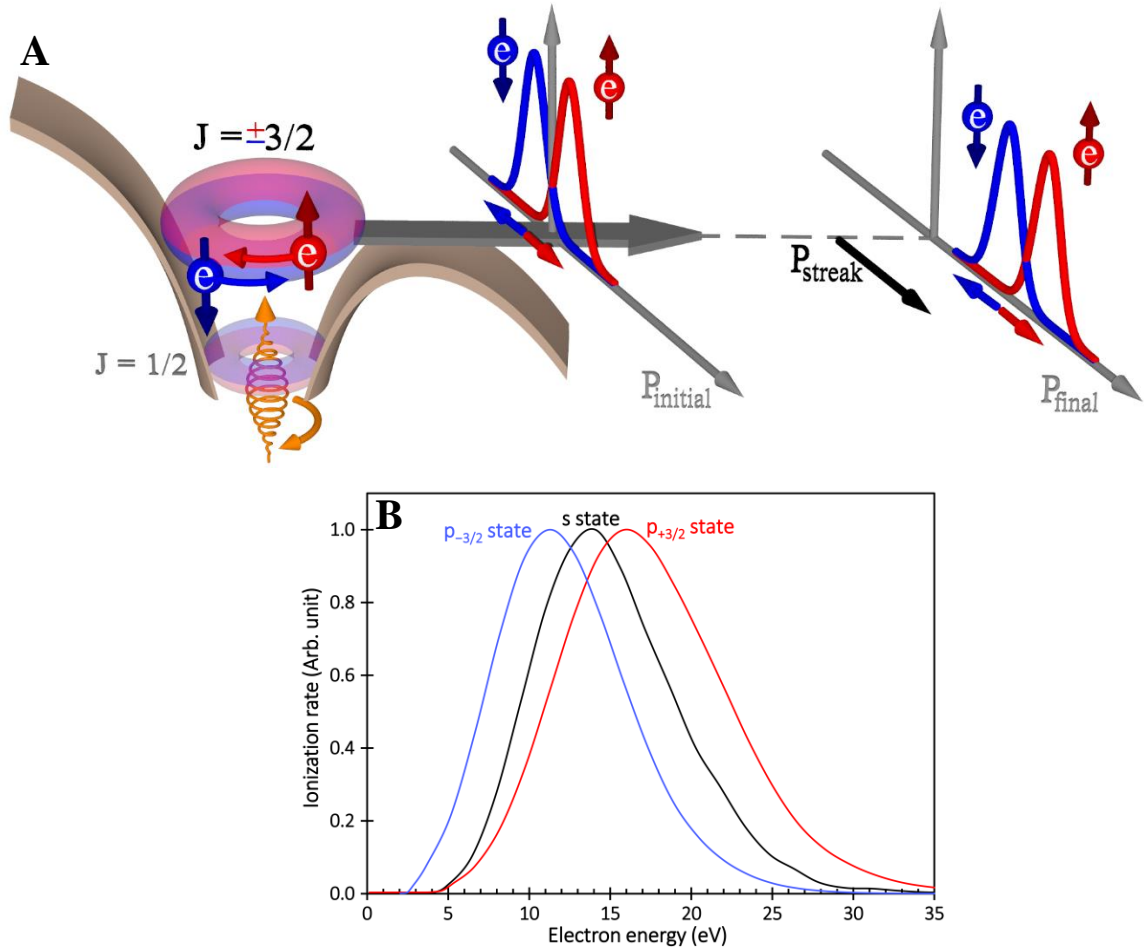

**Figure 2 | A: artist view of ionization process of the Xe 5p-state.** The 5p j = 3/2 states of Xe are predominantly ionized. In j = 3/2, |$m_j$|= 3/2 states, the electron angular momentum and its spin are parallel. The initial rotational state of the electron results in an offset momentum in the direction of the streaking momentum imparted on the electron by the laser field. Therefore, different spin states correlate to the different, shifted energy distributions, leading to energy dependent spin polarization. **B: theoretical energy distribution of s- and p-states.** The intuitive picture in panel A is confirmed by results obtained from the numerical solution of the time-dependent Schrödinger equation, for single active electron in the initial p-state. The effective potential for the electron motion is chosen to fit the 5p j=3/2 ionization potential of Xe. For the initial s-state, this potential was modified to maintain the same binding energy as for the p-state. The heights of the three distributions were normalized to unity.

This intuitive picture is supported by the numerical solution of the three dimensional time-depend Schrödinger equation. Figure 2B shows electron energy distributions $m_\ell$ = -1$\hbar$ (blue) and $m_\ell$ = +1$\hbar$ (red) together with the distribution for an initial s-state with of the same binding energy (12.13 eV). For the s-state, the distribution peaks close to $U_p$. The initial orbital momentum shifts that distribution, depending on whether the electron in the initial state co- or counter-rotates with the driving laser field.

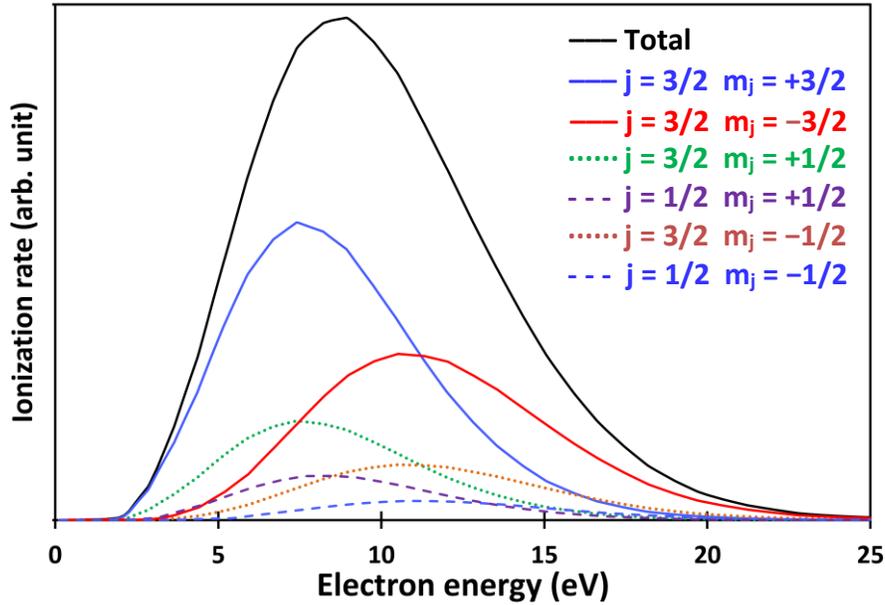

**Figure 3 | calculated kinetic energy distribution of electrons for different states.** States with different $m_\ell$ are offset in energy. $j = 3/2$ states have higher ionization rates than $j = 1/2$, because of the lower binding energy due to spin-orbit interaction. $|m_j| = 3/2$ states have higher rates than $|m_j| = 1/2$ due to the Clebsch-Gordan-coefficients. The field intensity is $I = 1.4 \cdot 10^{14}$ W/cm².

The real situation for Xe is somewhat more complicated than discussed so far. Firstly, also the $j = 1/2$ state contributes to the ionization. Due to spin-orbit interaction, in Xe the 5p $j = 1/2$ state is 1.3 eV stronger bound than the 5p $j = 3/2$ state of Xe ($I_P = 12.13$ eV) and therefore is significantly less likely to ionize (see figure 3). Secondly, in the $j = 3/2$ state not only $|m_j| = 3/2$ but also $|m_j| = 1/2$ contributes to ionization. For $|m_j| = 1/2$ for the same $m_\ell$ the orientation of the spin is opposite to the one correlated to $|m_j| = 3/2$. Fortunately, the addition of angular momenta described by the Clebsch-Gordan coefficients shows that the $|m_j| = 1/2$ state is three times less likely to be formed than the $|m_j| = 3/2$ state. Fig. 3 shows full energy- and channel-resolved results of our calculations as well as the total electron spectrum. The channel resolved spectra are then used to calculate the spin polarization as in Ref. 1, yielding good agreement with the experiment, see the solid curve in figure 1. The field intensity for the theoretical calculations in Fig. 3 was chosen such that the spin-integrated electron energy spectra for experiment and theory overlap each other, yielding the effective intensity of $I_{eff} \approx 1.4 \cdot 10^{14}$ W/cm² for the maximum of the electron distribution at 8.5 eV.

Experiment and theory both show a zero crossing of the spin polarization near the maximum of the peak in figure 3. This observation follows directly from the mechanism discussed above. When the electron does not have to traverse a thick barrier to become free, the initial transverse electron velocity distribution upon ionization is approximately a Gaussian, centered at 0, for an initial s-state. This Gaussian distribution is shifted symmetrically by the initial positive ($p^+$) or negative ($p^-$) orbital momentum (see figure 2). Therefore the ionization rates of spin-up und spin-down electrons have equal values at the center of the unshifted distribution (near $U_P$) and so the zero crossing of spin polarization is also at that position. We note that this is different from the predictions based on a spin polarization mechanism operating purely in the tunneling regime and illustrated for a short range potential where the barrier cannot be suppressed[1]. Tunneling through it in a circularly polarized field will lead to an initial momentum distribution after tunneling shifted away from zero, even for an s-state. As a result, all energy distributions and the zero crossing of spin polarization are shifted to higher energies ($U_P+I_P$ in this regime). Thus, zero crossing of spin polarization is a sensitive measure of the kinematics of the ionization process.

In conclusion, the significant degree of spin polarization in electrons ejected from Xenon atoms by a strong ultrashort laser pulse opens exciting new directions for strong-field physics. Extension of our results to single- and multi-color chiral laser fields supporting recollision[13,14] would bring the new dimension of the spin variable to laser induced diffraction[15], holography[16] and higher harmonic generation[17]. It would allow one to test chiral molecular systems with sub-femtosecond temporal and sub-Ångström spatial resolution. Our results show that orbital imaging can be extended to probe stationary and dynamical currents, e.g. in molecular orbitals. Application of modern few cycle circularly polarized pulses[18] would allow for production of sub-femtosecond spin polarized electron pulses, which then can be used to probe magnetic properties of matter of ultrafast time scales[19]. Finally, spin polarization of the ejected electron is firmly linked to the creation of the parent ion in an initially spin polarized state. Spin-orbit coupling then leads to an internal circular electron and spin current, confirming recent predictions of Ref. 2.

## Methods

### Experimental measurements

For the measurements a commercial KMLabs Dragon Ti:Sa laser system (40 fs, 780 nm, 0.5 mJ per pulse) was employed. We used a quarter-wave plate to produce circular polarization from the initially linearly polarized light. The ellipticity of the electric field for left circular polarization of 0.96 (for right circularly polarized light = 0.93) was measured with a Glan polarizer and a rotational stage. The laser pulses were focused into the Xe gas target by a lens of f = 10 cm. The emitted electrons travelled through a 50 cm field-free drift-tube to a commercially available Mott-Detector[12], which is capable of measuring the spin polarization of an electron beam. Due to spin-orbit interaction the differential cross-section of electrons scattered at High-Z atoms (in our detector a thorium-target is used) is spin-dependent. A Mott-Detector utilizes this effect and measures the scattering-asymmetry A of electrons. A is given by $(N_U - N_D) / (N_U + N_D)$, where $N_{U,D}$ are the number of electrons scattered upwards respectively downwards. A is related to the spin polarization P by $A = S_{eff} \cdot P$, where $S_{eff}$ is a constant of proportionality defined by the detector geometry. For the instrumental scaling factor $S_{eff}$ of our Mott polarimeter we used $S_{eff} = -0.15$. The literature value is between −0.15 and −0.25 for an acceleration voltage of 25 kV. In the measurements 18 kV were used and we therefore reduced $S_{eff}$ accordingly[12]. Additionally to statistically measuring the polarization we gained information on the kinetic energy of each electron by recording its time-of-flight. Instrumental imperfections of our setup, e.g. different detection efficiencies, would lead to an asymmetry, which is indistinguishable from the measured asymmetry caused by polarization. To cancel those, we made two measurements with left and right circularly polarized light. Between those measurements the polarization effect should just switch sign whereas the instrumental asymmetry stays the same. For comparing the two measurements in analysis, a third MCP detector was employed, lying in the plane of laser propagation and hence being unaffected by a possible spin polarization in that axis. Small intensity differences between the measurements with left and right circularly polarized light were leveled out by comparing the energy distributions of this third detector. The distributions were put on top of each other by stretching one of them and assigning this stretch factor to the other two spin measuring detectors.

**Numerical simulations**

For the theoretical calculations in Fig.2, we have numerically solved the time-dependent Schrödinger equation, using the single active electron approximation, with the electron in the initial p$^+$ or p$^-$ state, i.e. co-rotating or counter-rotating with the laser field. We have used the effective model potential[20] $V(r) = -\frac{1}{r} - \frac{(Z-1) \cdot e^{-\kappa \cdot r}}{r}$ with Z=54 and the parameter $\kappa = 1.2285$ adjusted to fit the lowest ionization potential of the Xe 5p shell, $I_p$ = 12.13 eV. The electron spectrum was obtained by propagating the wave function for additional 2 cycles after the end of the laser pulse, during which electrons with e.g. 9 eV energy move by an additional 170 a.u. away from the origin. Next, we extracted the continuum part of the wave function using a spatial square mask with the radius 100 a.u. which eliminates the central part of the wave function near the core. This masking procedure is adequate, because for an intense circularly polarized pulse the electronic wave function is well separated into the bound part near the core and the continuum part far away from the core. The remaining continuum part was then projected on the plane wave basis. The accuracy of this procedure has been monitored by varying the additional propagation time up to 5 cycles, varying the radius and the width of the spatial mask, and comparing the spectra obtained with the same laser conditions and using the same procedure, but for a hydrogen atom, against the exact spectra obtained by projecting on the well-known exact continuum eigenstates of hydrogen. For the initial s-state (in Fig 2), the effective potential was modified to maintain the same ionization potential binding energy as for the two p-states. We have used different pulse shapes to test the validity of the numerical analysis: 2 cycles sin² ramp up and 2 cycles sin² rump down, 4 cycles sin² ramp up and 4 cycles sin² ramp down, and a "long" pulse with 2 cycles sin² ramp up and down and 4 cycles flat top. Apart from angular asymmetry introduced by the two shorter pulses, the angle-integrated electron energy distributions and spin polarization remained essentially the same. All the numerical data shown in the figures are for the "long" pulse. For the calculations in Fig. 2 the amplitude of the circularly polarized electric field was set to F = 0.05 a.u. For the spin-orbit channel resolved calculations in Fig.3 we have modified the effective potential, to fit the ionization potential of the 5p j=1/2 channel ($I_p$ = 13.44 eV) following the recipe described in Ref. 21. Specifically, an additional short range potential step was added to the effective potential. The step is non-zero only at the first grid point, fixed at 0.5 a.u. The field intensity $1.4 \cdot 10^{14}$ W/cm$^2$ was chosen such that the spin-integrated electron energy spectra for experiment and theory overlap each other. We have then followed the prescription in Ref. 1 to compute spin-polarization depicted in Fig.1, performing additional averaging over intensities I = 1.1 - 1.4 $\cdot$ 10$^{14}$ W/cm$^2$.

**Acknowledgements**

The experimental work was supported by the Deutsche Forschungsgemeinschaft. A.H. and K.H. acknowledge support by the German National Merit Foundation. M.I. acknowledges support of the EPSRC Programme Grant EP/I032517/1 and the United States Air Force Office of Scientific Research program No. FA9550-12-1-0482. F.M. and O.S. acknowledge support of the DFG grant SM 292/2-3.

**Author contributions**

All authors edited and commented on the manuscript. A.H. and M.K. carried out the measurements. A.H., M.K., K.H., M.R., T.J., A.K., M.S. and L.S. built up the experimental setup. A.H. and A.L. analyzed experimental data. F.M., M.I. and O.S. carried out the numerical simulations. R.D. supervised all work.

**Additional information**

The authors declare no competing financial interests. Readers are welcome to comment on the online version of the paper. Correspondence and requests for materials should be addressed to A.H. (hartung@atom.uni-frankfurt.de).